\documentclass[prl,onecolumn]{revtex4}
\usepackage{txfonts}
\usepackage{graphicx}
\usepackage{natbib}   
\usepackage{amssymb}

\begin{document}

\title{Dark Matter Strikes Back}

\author{Paolo Salucci} 
 
\affiliation{ SISSA, Via Bonomea 265, 34012 Trieste, Italy}

\date{\today}

\pacs{95.30.Sf,  95.35.+d, 98.52.Nr, 98.52.Wz, 04.50.Kd}

\begin{abstract}
Mc Gaugh et al. (2016) have found, by investigating  a large sample of  Spirals,   a tight non linear relationship between  the  total radial  acceleration,
connected with the Dark Matter phenomenon,  and its component which  comes  from the distribution of  baryonic matter,  as  the stellar and HI disks. The strong  link between these two quantities   is considered by them  and  by other researchers, as challenging the scenario featuring the  presence of DM halos in  galaxies. Or, at least, to indicate the peculiar nature of the underlying dark matter particles.  We have explored this issue by investigating a larger number of galaxies  by means of  several techniques of analysis.  Our results  support and even  increase,   both qualitatively and quantitatively, the validity  of  McGaugh et al. (2016) 's relationship.  However,  we prove that such relationship exists  also  in the scenario featuring dark matter halos + ordinary baryonic matter  and  that it arises by the fact the DM is less concentrated than the luminous matter and  it is progressively more abundant in lower luminosity objects. These properties are due  to well known astrophysical effects: the  implications of this  relationship for the properties of dark matter halos are nothing of new or  of unexpected.  The relationship, definitively,  is not a  portal to go beyond the  standard picture of $\Lambda$CDM galaxy formation.

\end{abstract}
\maketitle
\section{Introduction} 
  
 Very recently,  a discovery seems to have blown up the fields of cosmology, astrophysics and elementary particles. The dark matter, the  elusive substance that  cosmologists  believe   to constitute  about the $25 \% $   of the mass energy of the  Universe and to  play a crucial role in the birth and  evolution of its structures,  seems to  have  disappeared. In the dark,  as  someone  is suggesting.  For the  past 30 years astrophysicists have believed  that spiral galaxies  were surrounded  by dark halos  (  \cite{rubin})  made by massive elementary  particles that  interact with the rest of the Universe  (almost) only by gravitation.  After a straightforward  analysis  of new  accurate data,  Mc Gaugh  et al. (2016)  claimed  that  this  scenario faces a challenge.  They found, from the  kinematical and  photometric data of  a 153  spirals,  that  their radial accelerations  show  an anomalous feature:  they  correlate, at any radius and in  any object, with their  components generated only  from the baryonic matter. This occurs   in a  very peculiar way which seems  to lack for  an immediate   physical explanation and to be at variance with the presence, around spirals,  of halos made by dark massive particles. Furthermore,  they claimed that, in any case,   this relationship  proves that  the  eventual  elementary particles  making the  dark halos  cannot collisionless.   

The aim of this paper is to show that, in the framework  of  Newtonian dark matter halos, the Mc Gaugh  et al. (2016) relation is just  a  low resolution realization of a  wellknown scenario which describes   the  {\it common}  past history of the dark and the luminous components of  Spirals.

In the present  framework  the  equation of centrifugal equilibrium reads :
\begin{equation}
 g(r)=g_{h}(r)+g_{b}(r)
\end{equation}
 $g, g_{h}, g_{b}$  are,   total radial acceleration  and its   components  generated by the DM halo and by the  baryonic matter, respectively.   
 In detail,  $g(r)=V^2(r)/r $, where $V(r)$ is the circular velocity and 
 \begin{equation}
 g_b(r) =(V_{D}^{2}+V_{B}^{2}+V_{HI}^{2})/r , \ \ \ \ \   g_h(r)=V_{H}^{2}/r 
 \end{equation}
 the  velocity fields $V_i$  are the solutions of the four separated Poisson Equations for the   dark and  baryonic components:
 $V^2_i= R  d\Phi_i/dR$ and $\nabla^2  \Phi_i= 4 \pi G \rho_i$ , where $\rho_i $ are dark matter, stellar disk, bulge, HI disk surface/volume densities)  ( $\rho_H, \  \rho_{B}, \ \mu_{D}(r) \delta (z), \  \mu_{HI}(r) \delta (z) $ ) with $\delta(z)$ the Kroenedeker function,   $z$  the cylindrical coordinate and $\Phi_i$  the gravitational potential.   Of course,  at any radius:  $V^2 =  (V_{D}^{2}+V_{B}^{2}+V_{HI}^{2})+V_{H}^{2}$
 
In statistical studies we can assume that the   main component of the  luminous matter is   the  well known  thin stellar  disk with an exponential surface density   profile: 
\begin{equation}
\Sigma (R) = \Sigma_0 e^{-R/R_D}
\end{equation}
where  $\Sigma_0= (M_d/L) I_0$  is the central surface mass density, with  $(M_d/L)$ the  stellar disk mass to light ratio,   $I_0$ the central surface  brightness,  $L $ is the total luminosity in a specific band  \cite{freeman}.  We will  work in the I band that, in Spirals, well represents  the stellar disk surface density.  We define $R_{opt}$,  the radius encompassing 83 $\%$ of the total luminosity/mass in stars,   as the size of the stellar disk. We have  $R_{opt} =3.2 R_D$,   $R_{opt}$   is  a quantity describing both the surface density profile and  the stellar disk size. Observationally, we have, (e.g. \cite{Tonini}):
\begin{equation}
\log \left( \frac{R_D}{\mbox{kpc}}\right) = 0.633+0.379\log\left( \frac{M_D}{10^{11}M_{\odot}}\right)   
+ 0.069\left(\log \frac{M_D}{10^{11}M_{\odot}} \right)^2,
\end{equation} 
that links  the spiral's stellar disk masses and their  sizes. 

We remind that, in  early  Hubble-type spirals, at $0 \leq r< 0.2 R_{opt}$,  a central bulge dominates  the  circular velocity while,  in the  low luminosity, late Hubble types spirals,  for  $r>R_{opt}$,  the  gaseus HI disk gives  a contribution to the circular velocity which is negligible with respect to that of the  dark halo,  but comparable with  that of the stellar disk.

The converging outcome   of two independent  lines of investigation evidentiates  that the dark and luminous  mass distributions in spiral galaxies  are  universal  specific  functions of  the disk mass $M_D$, (or of the disk Luminosity), and of the disk size $R_{opt}$ \cite{pss,ys}. Thus, with the help of   Eq(3)  and  with  the  disk mass (or the Luminosity) as the running variable in these functions, one obtains, for the entire family  of  spirals, the  dark and the  luminous matter  structural parameters  that straightforwardly lead to   the corresponding $g_h$ and $g$ accelerations.  Then, we  can carefully investigate, with thousands of objects,  the validity of the  McGaugh relationship,  whether it is compatible with the the dark matter halo scenario and, in such case, how weird is  the relative DM particle.

Let us stress that the relationships  in  McGaugh et al. (2016) and in  this work  regard  Spirals, in  Ellipticals and in  Dwarf spheroidals,  ,  the determination  of  the above  accelerations is very difficult due to  their complex  kinematics and dynamics and to their  poorly  known  DM halo density profiles (see e.g. \cite{fa}), thus no safe relationship can be established.

 In this work  velocities are in km/s, and distances in kpc accelerations $m/s^2$ 
 
 We proceed as it follows. In the second section we describe the Mc Gaugh et al. (2016) relationship, in the sections from the 3  to the 5 we work out such relationship by means of three different methods applied to 5 samples. The Implications and a  discussion of the findings are put forward in sections 6 and 7.  
Let us notice that  the Appendices A-C  are a resume' of  the  results obtained in previous works  that are  used here.   
  
\section{The Mc Gaugh et al. (2016)  relationship} 
 
 McGaugh et al. (2016)  investigated a sample of 153 galaxies across a large range of scales in luminosity  and Hubble Types and having  high quality (SPARC) rotation curves  $V(r)$.  
 For each object  they derived  the  total radial acceleration  $g(r)$ that  compared   with the corresponding value  of  the gravitational acceleration $g_b(r)$ generated, at the same  radius $r$ , by all the luminous matter of the galaxy.   McGaugh et al. (2016)   used the  surface brightness   and  the color of   each galaxy to  derive their stellar surface/volume mass densities.  These were   inserted  into  the  relative Poisson equation to provide them   with the  $g_b(r)$ values   out to  $   (1 -1.5)\  R_{opt} $, according to the  extension of each  RC.  The  systematics in  method used  is, at most 0.1 dex, and it has no effect on  the relationship found.  The entire  procedure is  sound: in this work,  the  McGaugh et al. (2016) determinations of  $g_b(r)$ and $g(r) $  are not  under discussion. The 1-sigma region of the Mc Gaugh et al. (2016)  relationship  is covered  in Fig (1) by  blue empty  circles. 
 
 The strong  correlation between the two acceleration  reads  as: 

\begin{equation}
g(r) =g_b(r)/(1-Exp[-(g_b(r)/a_0)^{0.5}])
\end{equation}
 with $a_0=1.2 \times 10^{-10} \ m/s^2$.
 
At low $g, g_b$ accelerations,   this relation is clearly very different from   that we  would expect  in the Newtonian no Dark Matter framework: $g(r)=g_b(r)$.  In fact, in this case,  the  centrifugal acceleration  must  perfectly balance  the gravitational acceleration due to the  whole  distribution of all the luminous  baryons  in the galaxy.   

This result may pave the way for  alternatives to the Dark Matter or  to the Galileian Inertia Law eg \cite{mo}. However, in this paper  we  will focus on  the Mc Gaugh et al. (2016)  strong  claim  that the relationship in  Eq.(5) is a  game changer also  within  the Newtonian  DM scenario. Thus,  the  issue in which  we are interested  concerns   the assumption of  the presence of dark matter halos around galaxies:  does  the above  two-accelerations relationship challenges it ? Does it  lead to a new understanding of internal dynamics of galaxies? Is it  tantamount to the existence of  elementary particles not  yet considered as candidate for Dark Matter ?  We will answer   these questions in the next sections.

\begin{figure}
 
\center\includegraphics[width=11.5cm]{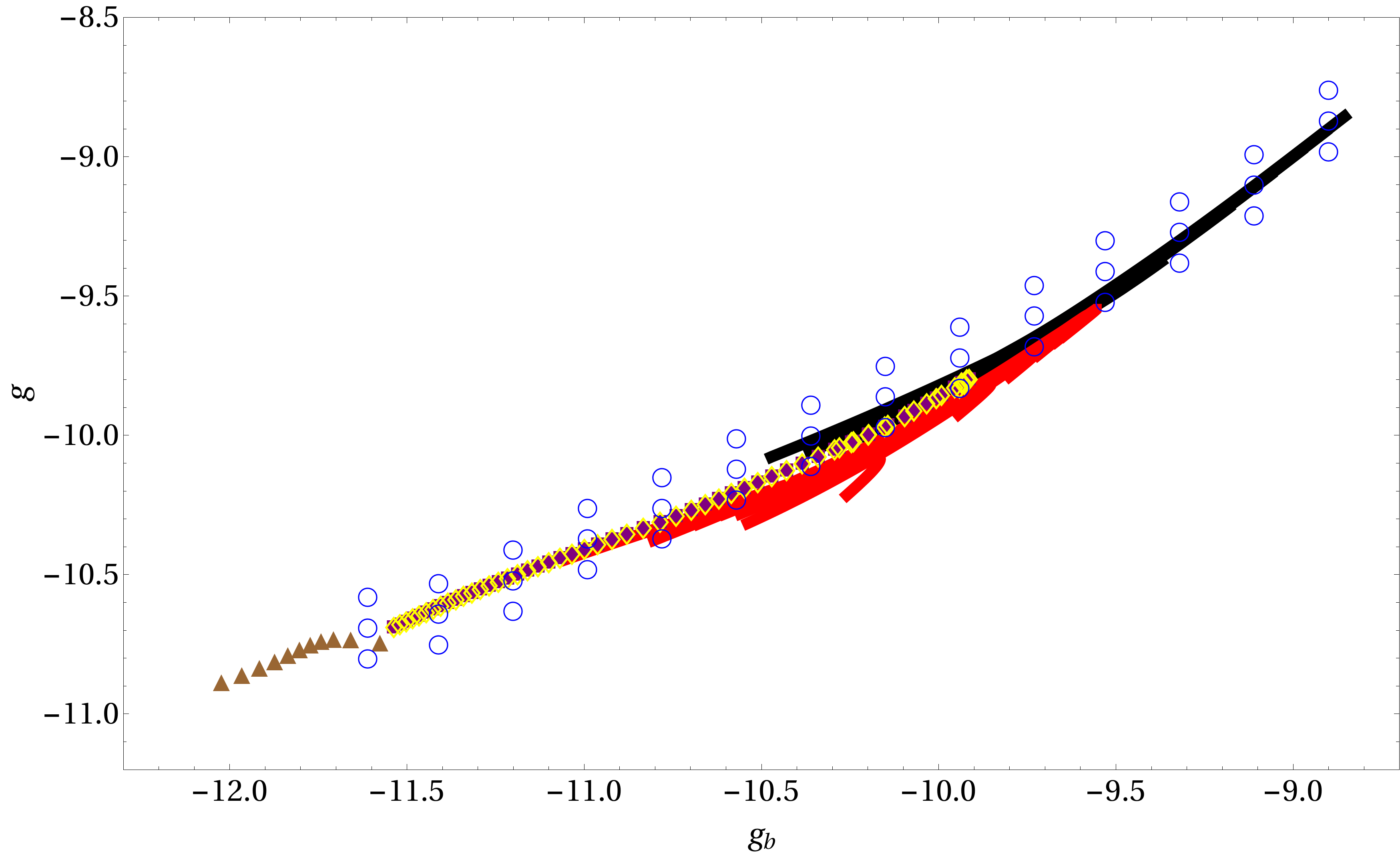}
\caption{The Mc Gaugh et al.  relationship and its 1-$\sigma$ uncertainties (blue circles), compared to  the relationships  obtained  by means of  the  URC  (red lines) and  the RTF methods (black lines), in M 33 for  Burkert and NFW halos (blu squares and yellow diamonds) and in  NGC 3741 (brown triangles).}
\end{figure}
  
\section{  The McGaugh et al. (2016)  relation relation from the Universal Rotation Curve }

We can  represent  the rotation curves (RCs) of late types Spirals by means of the Universal Rotation Curve (URC) pioneered in  \cite{rubin2,p91}  and set by \cite{pss}   and \cite{s7}.   
In short,  by adopting the normalized radial coordinate   $x\equiv r/R_{opt}$,    the RCs of Spirals  are well described by  a Universal  function of  $x$ and of $M_I$,   their absolute I magnitude. This function, $V_{coadd}(x, M_I )$,   has been  derived  by means of   the analysis of about  1000 rotation curves  (see Fig 1 of \cite{pss} )  and supported by following  investigations (see e.g.  \cite{ma})). We can  uniquely and successfully  fit   the   $V_{coadd} $ data by means of  a universal  analytical velocity  model   $V_{URC} (x, M_D ) $ which includes the following  velocity components: the standard Freeman  disk  $V_{URCD} (x, M_D )$,   a cored   dark matter  halo   $V_{URCH}  (x, M_D )$  and a  HI disk  $V_{URCHI}  (x, M_D )$ \cite {SB}(see Appendix A for details).    

Then, by means of the URC velocity model we  can  compute the above accelerations in a very large amount of  objects, 5.2 times larger that of  Mc Gaugh et al. (2016),  and at radii congruent with those of  McGaugh et al. (2016).  We have,  by writing now  the accelerations  explicitly as a function of $r$: $g_{URC}(r)=V_{URC}^2(x, M_D)/r  $.    The baryonic component of the  radial acceleration  is then  obtained  by removing from the latter  the Dark Matter halo contribution $g_{hURC}(r) =V_{URCH}^2(x, M_D)/r$,  so:  $g_{bURC}(r) = g_{URC}(r)-g_{hURC}(r) $.  

We  plot in Fig (1)  $ g_{URC}(g_{bURC})$  for objects in the above  I magnitude  range  and  for radii  $0 \leq r \leq 1.5\  R_{opt}$ (red lines  whose width indicates  the fitting uncertainty).

\section{The McGaugh et al. (2016)  relation relation from the Radial Tully Fisher relationship }
 
Yegorova and Salucci (2007), by analyzing three different samples  of 794, 86 and 91 spirals of different Hubble types that  include  a fair  number of  objects with a prominent bulge,  discovered the  Radial Tully-Fisher (RTF) relation (see \cite{ys}). This is  a series of tight and  independent relations  between   the galaxy  absolute  magnitude $M_I$ and   $\log V_n$,  the circular velocity measured, in each object, at  the same fixed fraction $R_n/R_{opt}$ of  the disk size $R_{opt}$. In detail,  $R_n= (n /10) \ R_{opt}$, $ 1\leq n \leq 10$, ($R_{opt} \equiv  3.2 R_D$). In detail: 
\begin{equation}
M_{I} = a_n \log V_n + b_n\,,
\end{equation}

with   $a_n$ and $b_n$  the parameters of the fits and  $n$ the indicator of the  radial coordinate.  Due to the limited number of optical  kinematical  data in the  outermost regions of spirals, the RTF  relationship is established  only out to $R_{opt}$.   All  the  10 relationships in  Eq. (6)   are independent and  statistically relevant   \cite{ys}. 
The values of the $a_n$  parameters are very similar in all the  three  samples  and we have \cite{ys} :  
\begin{equation}
a_n=-2.3-9.9 (R_n/R_{opt})+3.9 (R_n/R_{opt})^2 
 \end{equation}
 
The  strong  decrease  of $a_n$  with $n $ in Eq. (7)  implies that  in Spirals  the stellar  light does not follow the distribution of the  gravitating  mass,   which would  require: $a_n=cost \ \ \simeq -7.5$.  As consequences,  a) we must insert in the  circular velocity model   also a  dark component, alongside the  baryonic  components a   b) all velocity components  must  be function of  $(x, M_I)$ \cite{ys} (see Appendix B for details).  Let us notice that  the velocity  model has  also a bulge component and the assumed  DM density profile can account  for cored  and cusped DM halos.  
By best fitting the outcome of the  RTF velocity  model to the function  $a_n(R_n)$  given in  Eq. (7)  we get the structural parameters of  its components and, in turn: $V_{RTFD} (r/R_{opt}, M_I ) $ , $V_{RTFB} (r/R_{opt}, M_I )$,  $V_{RTFH} (r/R_{opt}, M_I )$.  Noticeably,  the resulting best fit velocity model   prefers a  DM cored distribution, but, differently from the results of the URC  method, also cuspy models  lay  within the fitting uncertainties.  
(see Appendix B for details).

Then, by means of Eq. (4) we  can  compute the above accelerations in three samples with  large amount of  objects,   (4, 0.75,0.70) times wider,  than that of  Mc Gaugh et al. (2016)  et. , and at radii congruent with those of  McGaugh et al. (2016).   In detail:  $g_{RTF}(r)=V_{RTF}^2(r) /r   $ and $g_{bRTF}(r) =(V_{RTF}^2(r)-V_{RTFH}^2(r))/r$.   

We  plot in Fig. (1)  $ g_{RFT}(g_{bRFT}) $ (black lines) by running the  absolute magnitude $M_I$  within the  range   specified above and for  radii  between   $0.1\  R_{opt}$ and  $ R_{opt}$ .

\section{The Mc Gaugh et al. (2016)  relation  from the accurate  mass distribution of special objects  }

In this section  we work out the $g_b$ vs $g$  relation  for two special, test-case spirals.  We will  reproduce their extended high quality  RCs   by means of a  Individual Rotation Curve  velocity model  which includes a  Freeman stellar disk,  a   HI disk    and  a dark halo.  We will derive  $V_{IRCH}$ out to large galactocentric distances.
 
M33 is a low-luminosity  galaxy  important for  investigating the distribution of dark matter in galaxies.  In fact,  it has  no bulge, it is very rich in gas and,   due to its proximity, its  rotation curve  has an excellent spatial resolution \cite{cs}. At outer radii, the HI  disk is  the major  baryonic   contribution to the  galaxy circular velocity.       Moreover,   M33 is  one of the fewest objects  whose rotation curve is well reproduced both by  a NFW cuspy  and a  cored DM halo density profile \cite{cs}.  

NGC 3741 is a dwarf spiral  with  the most extended available  HI disk rotation curve  in terms of the galaxy stellar disk size lenght scale,  $R_D$ \cite{gs5}.  This galaxy has been observed in the HI 21cm  line with the Westerbork Synthesis Radio Telescope out to  42 $R_D$ i.e.  $13.5 \ R_{opt}$.   The  rotation curve, accurately derived from  HI  data cubes,  is decomposed into its stellar, gaseous and dark components,  the latter represented by a Burkert  halo, necessary, in this object,  for  a successful fit. 
The  best fit   parameters  for the IRC velocity model are shown in Appendix C.     

Then,   for these  top cases  of  the investigation of dark matter in spirals,  we directly  obtain  the radial accelerations $g(r)=V^2_{IRC}(r)/r$, while their  baryonic contributions $g_b(r) $ are obtained   by means of  the  differences  between  the radial accelerations and  their Dark Matter components:    $g_b(r)=  (V^2_{IRC}(r)-  V^2_{IRCH}(r) )/r   $.

Then, for the 2 objects,   we get: $g=g_{IRC}(g_{bIRC})$ that we plot in figure (1) (as points).

\section{Discussion}
 
 In this work,  within the Newtonian  Gravity, we have   investigated the   $g_b$ -  $g$  relationship by  McGaugh et al. (2016) in  several  large samples of  rotation curves of spirals,  spanning a very wide  range  in  luminosities and  Hubble types.   In an approach complementary  to theirs and with about ten times data than them,  we have  {\it assumed}   the presence  of  dark matter halos  as the origin  of  the ``anomalies''  of spiral's kinematics in particular the  relationship under study.  
 In detail, by means of proper  velocity  models  including, in all cases,    a  stellar disk  and a  DM  halo, in two cases also a  HI disk, and  in one case also a  stellar bulge, 
 we have derived, by means of  three  different methods, the above  accelerations.
  The first method  we used comes from the phenomenology of spiral rotation curves and  the inability of  the luminous matter  to match  the  circular velocity profiles  without the assistance of  additional  specific  dark halo component.  This method  is  very sensitive to the distribution  of the DM inside $R_{opt}$. 
The second method  exploits the  existence,  at different normalized radii of  spirals,  of tight   kinematics vs luminosity relationships, which yields the  DM halo velocity profile.  This method considers  the stellar  bulge component   and it is very sensitive to  the {\it amount}  of  DM  inside $R_{opt}$ . 
The third method is the standard  mass modelling  of  extended  high resolution rotation curves  with optical   and HI photometry which provides us with DM halo velocity profile.   This method is  sensitive  also   to the distribution of baryons well outside $R_{opt}$  

These 3 different methods were applied to  5 Samples for an amount of estimated  (15500, 6500, 1500, 1500, 100) independent data, very much deepening  the investigation by  McGaugh et al. (2016)  et al.

In all the cases,  $g_b(r)$, the     baryonic component of the radial acceleration  is obtained by subtracting  the  DM halo contribution from  the total  radial acceleration $g_b(r)=g(r)-g_h(r)$.  This component  is so derived in a different and even opposite way with respect to  that in  Mc Gaugh et al. (2016)  et al where, instead,  it   is obtained {\it directly}  from the properties of the luminous matter  with no involvement of the  Dark Matter ones. 

All our  methods  applied to our samples leads to a same  universal $g=g(g_b)$ relationship, see Fig. (1).  Its comparison  with that found by  Mc Gaugh et al. (2016) et al   has  striking implications.   The  $g(r)$  and $g_h(r)$ accelerations obtained by means of  our  velocity models in our samples   lead to   essentially  the same  relationship of  Mc Gaugh et al. (2016),  see Fig. (1).  Let us stress again that our framework has  a DM halo in every galaxy whose presence  plays the  crucial role  of  bending  the  expected  $g_b=g$ relation into one which results very consistent with that found by  McGaugh et al. (2016) (see Fig. (1)) which is, instead, obtained in a way completely agnostic to the DM  presence. 
  
Therefore, considering the very good agreement of these two approaches,   there are no reason  for  claiming  that the Mc Gaugh et al. (2016) relation  is alien to the scenario featuring  DM halos  in Galaxies or that   it  challenges it. On  the contrary, any  scenario which does not feature such  dark halos  is obliged to   interpret  our mirror relationships  as  repeated   coincidences.  Furthermore, in the present  dark matter halos framework, the   Mc Gaugh et al. (2016)  relationship unravels no secrets  about  the dark halo component in Spirals. Specifically,  it yields no relevant  information about  the DM halo  density   profile:  in fact,  we find that   velocity models both with  NFW and  with a cored DM profile  lay  within   the  uncertainties of   the Mc Gaugh et al.  relationship.  The intrinsic  nature of the  dark particle is beyond  the physics underlined by such relationship .
\begin{figure}
\includegraphics[width=11.5cm]{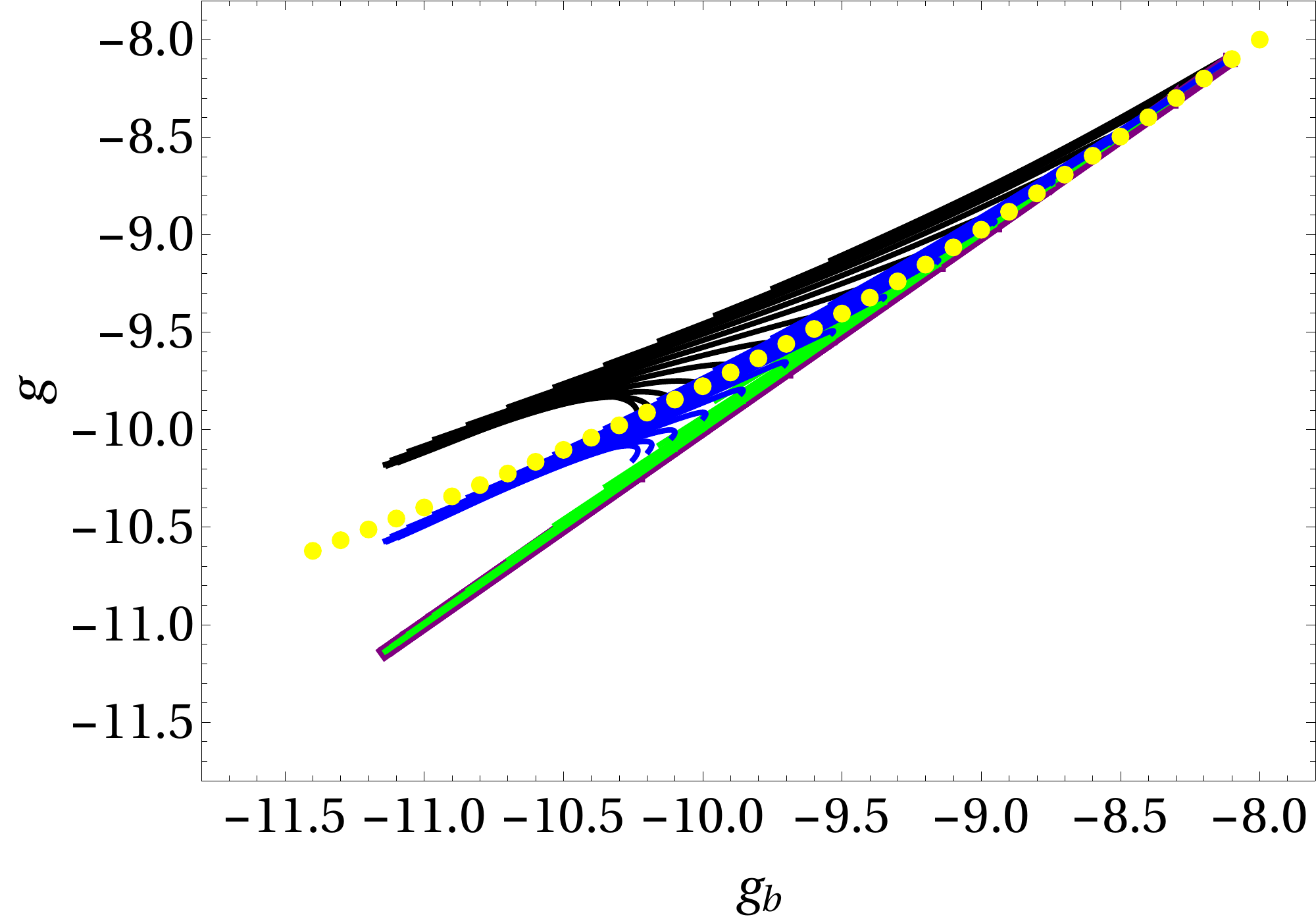}
\caption{The Mc Gaugh et al. (2016) relationship (yellow points) and  those found in this work (blue lines).  The relationships that  emerge, for the same luminosities and radii,  in the cases of  -no dark matter (red lines), -compact dark matter (purple lines)  - all dominating dark matter (black lines)  and   -increasing with luminosity  fraction of  DM(green lines),   are also shown.}
\end{figure}
 
The next step is to  explain  the Mc Gaugh et al. (2016) and ours relationship  in terms of well understood general properties of dark matter. Let us  assume a General velocity profile $V_G(x)$ and the  relation in Eq. (4).  For the stellar disk  component  $V_{GD}(x)$ , we adopt the usual  velocity profile of  Eq. (10), for the HI disk contribution  $V_{GHI}(x)$,  the approach described by Eq. (11).   Then, the dark halo component  reads:
\begin{equation}
 V_{GH}^2(x) = 2.65 \  10^5 M_D/R^2_D \   B \  x^d/(a+x^2) \ ( M_D/(10^{11}\ M_\odot) ^c  
\end{equation}
  where $B$ is proportional to the fractional content of dark matter at $R_{opt}$ , $ c $ indicates  the dependence of the latter quantity on  the disk mass,   $a$ indicates the size  of  DM halo core and  $d$ indicates how compact is the distribution of dark matter with respect to that of the stars.
  
The  General halo velocity model in Eq. (8)  includes the models adopted in the previous sections, but  it can represent  also very different ones.    Obviously,  at any $x \equiv r/R_{opt}$,  we have:  $ V_{G}^2= V_{GH}^2 +V_{GD}^2+ V_{GHI}^2$.   The disk mass is the running variable.  

We have: $g_G(r)= V_{G}^2/ r$  and: $g_{bG}(r)=(V_{G}^2-V_{GH}^2 - V_{GHI}^2)/r$. Now we invert the process performed in the  previous sections and we use the  McGaugh et al. (2016)  relationship  data  to  get the  four free  parameters of the General model of spiral velocities . We found:  $a=1$, $c=-1/2$, $d=2$, $B=0.1$, with  the formal  fitting uncertainties  on the parameters running from 20\% to 50\%.     Noticeably, these best fit values for the  parameters are very similar to those  we have assumed in the  the three  velocity models considered in the previous section (see Fig. (2)). This is expected in that the General model does include  our velocity models   that, in turn,   yield  the  Mc Gaugh et al. (2016)  relationship. 

Let us  investigate  the situations in which the General model relationship {\it  fails}  to reproduce the  Mc Gaugh et al. (2016)  relationship.  The quantity $a$ plays no role in the agreement between  these two relationships: we can take  $0.4 < a <\infty $ without breaking it.  Instead, for values of the quantity $B$  such as:   $B\simeq 0 $ (no dark matter) or $B>0.3$ ,  i.e.  for an  amount of  DM $ >  3$  times the best fit value,  the agreement breaks down and the General model fails to reproduce the Mc Gaugh et al. (2016) (and ours) relationship  (see Fig (2)). Similarly, the agreement continues also for values of  $d$   different from the best fit value of  2,  but,  for  $d<-2/3$,  i.e. for  a DM halo distributed in a  more compact way than the luminous matter,   the agreement breaks down (see Fig. (2)). 
 
Therefore, in the Newtonian dark matter scenario,  the Mc Gaugh et al. (2016) et relationship, enhanced by the results of this work, {\it exists}  since and only since 
a) the luminous matter is more concentrated than the dark matter: the quantity $g_h(r)/g_b(r)$ increases with radius.   
b) in  lower luminosity objects there is a larger fraction of dark matter: the quantity $g_h(R_{opt})/g_b(R_{opt})$ increases with decreasing galaxy luminosity.  

Notice that:

1) a) This evidence  was known  since the very discovery of DM in spirals  ( e.g.  \cite{rubin}) and it comes  from the most important property of  the  Dark  Particles:  they do interact  with baryons only gravitationally. It is generally agreed that in  protohalos  the dark matter and the baryons  were distributed in undifferentiated way,  but  during  the following  assembling of the stellar disks,  the infalling  baryons did  dissipate much of  their kinetical energy and fell deep inside  into the galaxy potential well.  The collisionless  DM particles instead,   conserved all their primordial  kinetic energy  and  populated  the outer parts of the halos. 
 
2) b)  This evidence is known since \cite {sal} and it  is easily  explained by  the fact that the lower is the  luminosity  of a galaxy  and then  so its gravitational potential well, the  more efficiently  the  energy injected into the interstellar space by Supernovae explosions has  removed  the  neutral hydrogen from  the galaxy,  preventing it to be  turned into stars. 

3) a) and b) clearly emerge in most of  related simulations and semi analytical studies  ever performed  in  $\Lambda CDM$ scenario. 

\section{Conclusions} 

There are strong   views that the  Mc Gaugh et al. (2016) relationship  is a very special one. Infact,  it connects  two physical quantities measured  at a same place in all objects and at all radii:  the relation comes out   independent of the galaxy  magnitude, color, maximum circular velocity, central brightness, Hubble type, stellar disk lenghtscale, HI content and present  star formation rate.  Moreover,  it allows an observer  who measures her/his radial acceleration with respect to the center of the galaxy $g(r)$,  to know,  at the same time,  the gravitational acceleration  $g_b(r)$  due to all  baryons of the galaxy s/he is subject to and this despite  the evidence  that, in galaxies, the light does not trace the gravitating mass.  The claims that we  are facing a meta-universal relationship, according to which, the distribution of dark matter is subjected to that of the luminous matter (or viceversa) seem justified. 

Instead, the results of the present  work imply that  such  view is  just a mirage. Although the Mc Gaugh 2016 relationship is connected to physics, in that  it arises  from a combination of physical effects:  the collisionless nature of the dark particle, the infall of baryons  in the DM halos   and  the  energy deposited  into  the interstellar space by  Supernovae explosions, however, it is not directly related to a specific physical process, least of all it leads to new physics  or to exotic particles.

Concluding, the presence of DM halos  in galaxies is perfectly compatible with the McGaugh et al. (2016)  relationship, that,  obviously,  cannot be considered as an evidence against the DM halos of elementary particles hypothesis.  The  McGaugh et al. (2016)   relationship  is  reproduced by  theoretical - numerical studies within the $\Lambda CDM$ scenario \cite{fa} but yields  no valuable  information  about the Dark  Matter distribution  in Spirals and the  related  intrinsic nature, differently from other galaxy  scaling laws  (e.g. see \cite{ma}).  Then,  in order to  open  the portal  to new physics,  we should head towards  other  evidences that dark and luminous matter behave  in galaxies as active  partners, and not as components which just share a common gravitational well, (e.g. \cite{D}, \cite{g} )

A 10-min video in which I will discuss this paper, avalaible at: https://www.youtube.com/watch?v=8K-VCoXJdus\&t=20s

\section{Appendix A}

$V_{coadd}$  is obtained from the PS sample of  967 galaxies in the following way:  1. We divide  the full spiral I  magnitude range:    $ -23.5 \lesssim M_I \lesssim -17$  in 11 successive  bins, each of them  centered at  $M_I$,  as listed  in \cite{pss}  We  assign each RC of the Sample  to its corresponding luminosity bin.  We coadd,  in normalized coordinates $ x=r/R_{opt}$,  all  the RCs assembled in each luminosity  bin,  and  we average them to get  $V_{coadd}(r/R_{opt}, M_I )/V_{coadd}(1, M_I )$ see Fig (1)  of  \cite{pss} 

We fit these   $V_{coadd}$ data with the   URC  velocity model defined as. 
 \begin{equation}
 V^2_{URC}(x, M_D)= V^2_{URCD}(x, M_D)+ V^2_{URCH}(x, M_D)+V^2_{URCHI}(x, M_D)
 \end{equation} 
the disk component is given by: 
\begin{equation}
 V_{URCD}^2=\frac{  G M_{D}}{2 \ R_{D} }  \left( 3.2x\right) ^{2}\left(I_{0}K_{0}-I_{1}K_{1}\right)  
\end{equation}
with the Bessel functions evaluated at $1.6 x$ .

 The HI  component  is modeled as \cite{e, Tonini}:  with an exponential thin disk of lenght scale $3 R_{D}$  
 \begin{equation}  
 V_{URCHI}^2= V_{URCD}(x)  \  10^7 (M_\odot /M_D)^{0.8} (1+( M_D/(3.3 \times 10^{10} M_\odot)^{0.75})
\end{equation}
with the Bessel functions evaluated at $1.6/3 x$ . 

For the dark  halo we assume the  Burkert profile density profile:  $\rho_{URCH}(r)=\frac{\rho_0 r_0^3}{(r+r_0)(r^2+r_0^2)}$  with   $r_0$ the core radius and $\rho_0$  the central halo density the free parameters, we have  

\begin{figure}
\centering
\includegraphics[width=9.5cm]{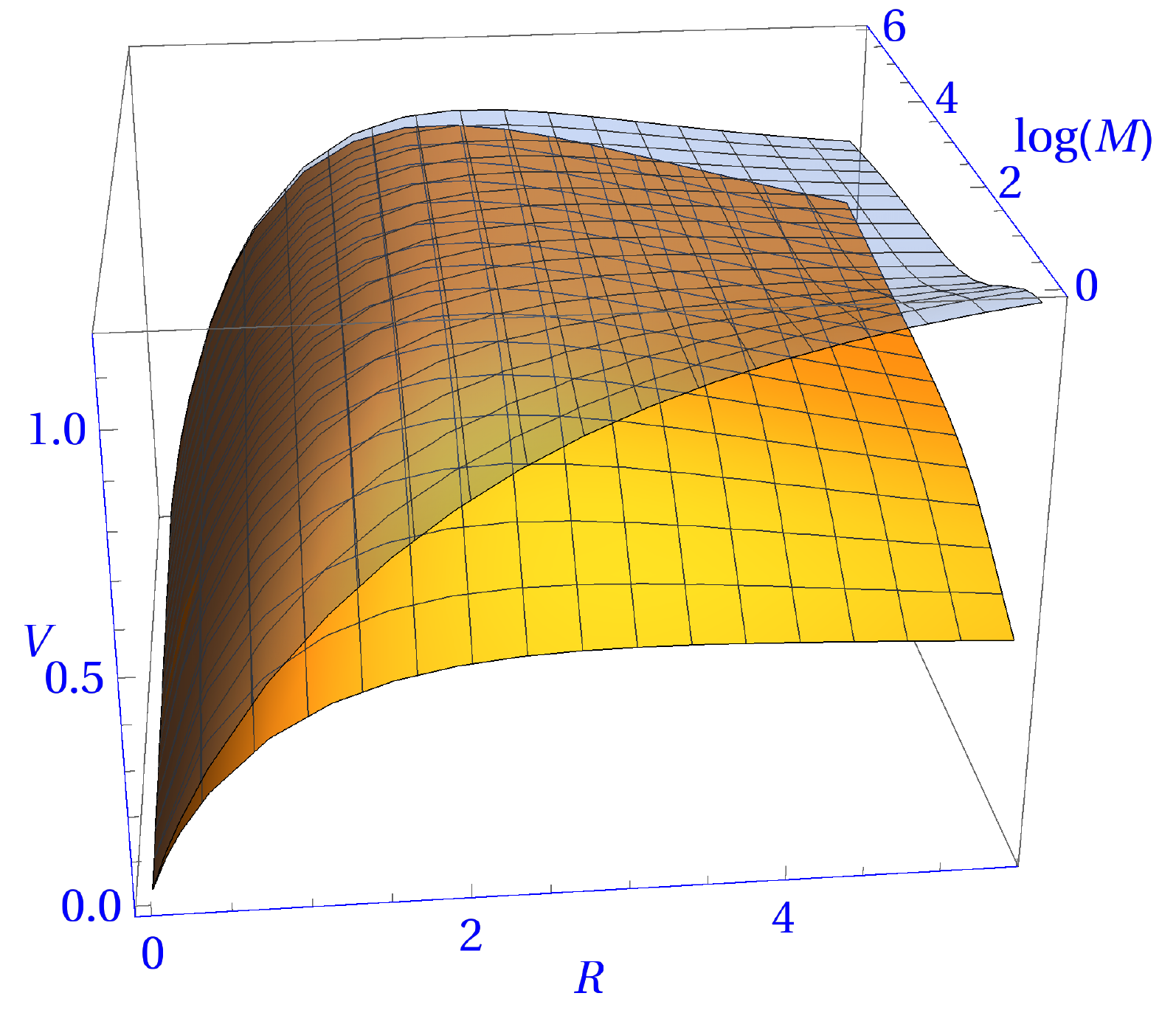}
\caption{$V(x, M)/V(1, M)/$ and $V_b(x, M)/V(1, M) $ obtained by the URC.   $ log \ (M_D/M_\odot)= 11 + 1/3 \ log M$  }
\end{figure}

\begin{equation}
V^2_{URCH}(r) =6.4\frac{\rho_0r_0^3}{r}( \ln (1+\frac{r}{r_0})-\arctan \frac{r}{r_0}) +\frac{1}{2}\ln ( 1+\frac{r^2} {r_0^2})).                       
\end{equation}
we get :
\begin{equation}
 log (M_D/M_\odot) =   -0.52 M_I-0.45 
\end{equation}
\begin{equation}
\log {\rho_0\over \mathrm{g}~\mathrm{cm}^{-3}} = -23.515-0.964\,
\left({M_D}\over 10^{11}\, \rm M_{\odot}\right)^{0.31}~   
\end{equation}
  \begin{equation}
\log \left( \frac{r_0}{\mbox{kpc}}\right) \simeq 0.66 +0.58 \log\left( \frac{M_{vir}}{10^{11}M_{\odot}}\right)
\end{equation}

that give us the URC dark halo  component:  $V_{URCH}(x, M_D)$ for  $ 0 \ \geq r/R_{opt} \leq 2$.   $V_{URC}$ and $(V_{URC}^2-V_{URCH}^2)^{1/2}$     obtained by this method  at any radius and for any disk mass are shown in Fig. (3).  We remind that = $g_i=V_i^2/r$ 
The uncertainty in the estimate of the various accelerations is about   $ < 20 \% $,  i.e. negligible  for  the aim of this paper. Notice that $ g_b(r)<< g(r)  $ in the region where the HI disk gives a contribution to $g_b(r)$ bigger than the disk contribution.   
 
\section{Appendix B}  

It is useful to define $l$ as the fraction of a spiral I-band  absolute luminosity with respect to  $M_I=-24$,  a reference  magnitude corresponding to  the most  luminous spiral in our samples,  $ l= 10^{-(M_I+24)/2.5}$.  The RTF velocity model has 3 components  (disk, bulge, halo) with: $V^2_{RTF}= V^2_{RTFD}+ V^2_{RTFH}+ V^2_{RTFB} $.
Let us notice  since the  RTF  method  uses  relationships  inside  $R_{opt}$,  we neglect in the velocity model  the (very small)  contribution to the circular velocity  due to the HI disk.

For the  stellar disk velocity  component $V_{RTFD}$ we assume the standard Freeman law: i.e. the RHS of eq (10)
For the bulge  we assume  the  Hernquist  mass profile with an  half mass radius $ 0.16 R_{opt}$. The bulge mass  is assumed to be a   fraction $c_B l^{0.5}$ of the disk mass $M_D$;  the power law index $0.5$ is suggested by the bulge-to disk vs total luminosity relation in spirals.  Then 
 
\begin{equation}
V^2_{RTFB}= \frac{1.21 c_B l^{1.3} x}{(x+0.1)^2}
\end{equation}

The amplitude  of the halo contribution to the circular  velocity  is related with that of the disk by  means of  2 parameters:  
$V^2_{RTFH} (R_{opt})=  c_h ( l)^{(k_h - 0.5)} \  V^2_{RFTD} (R_{opt}) $

For the halo  velocity contribution we adopt  the following profile: 
\begin{equation}
V^2_{RTFH }=c_h l^{(k_h - 0.5)}  ({x^2\over {x^2+\alpha^2}})(1+\alpha^2),
\end{equation}
 This profile, used  in  \cite{pss},  can represent both   Burkert halos,  in which   $\alpha  > 1$ and   NFW halos, in which    $\alpha  < 1/4$ 
The parameters  $c_h$ ,$ k_h $ , $c_b$  and $\alpha$  specify completely the velocity model.  We obtain them  by  best reproducing    the  $a_n=a(R_n)$  relationship. We found  $k_h=0.79 \pm 0.04$,   $c_b=0.13 \pm 0.03 $,  $c_h= 0.13 \pm 0.06$,   $\alpha = 1 ^{+1}_{0.5} $ in units of $R_{opt}$. The  uncertainties on the  LM  contributions to the circular velocity are  $ < 30 \% $,  i.e. negligible for  the aim of this paper.   $V_{RTF}$ and $(V_{RTF}^2-V_{RTFH}^2)^{1/2}$   obtained by the RTF method at any radius and for any luminosity  are shown in Fig. (4).  We remind that : $g_i=V_i^2/r$  
 
Let us stress:  1) The RTF method deals with the bulge component in spirals and it shows that this component, as we can predict, has its play  in the relationship at high accelerations when $g=g_b$   2) The best fit value  for $\alpha$ does not  well discriminate,  within its uncertainty, whether the DM density halos are cuspy or cored. Therefore the  $g_{RTF}(g_bRTF)$  relationship, plotted with its uncertainties in  Fig (1),   represents both cored and cusped DM density models.

 \section{Appendix C}  
 
For an individual rotation curve, at any radius: 
 \begin{equation}
  V^2_{IRC}= V^2_{IRCD}+ V^2_{IRCH}+ V^2_{IRCHI}
 \end{equation}

M 33 Notice that   its  stellar disk surface  density  has  been  derived from multi-band optical imaging and the  correlations between galaxy colors and stellar mass-to-light ratios ({\cite{cs}). The structural  best fit values are in the NFW halo case :  concentration: $c= 9.5 \pm 1.5$, halo virial mass:  $ (4.3 \pm 1.0) \times 10^{11}\  M_\odot$ and  disk mass:  $(4.8 \pm 0.6) \times  10^9  M_\odot$  and, in the URC halo case:  core radius:  $r_0=(7.5 \pm 1.5)  kpc$, central density:  $\rho_0=(1.8  \pm 0.3) \times  10^{-2} M_\odot/pc^3$ and  disk mass  $ M_D=(7.3 \pm 0.6)\times 10^9  M_\odot$.  

NGC 3741 The structural best fit values are: $r_0=(3 \pm  0.5) kpc$,  $\rho_0=(1.6\pm 0.3) \times  10^{-24} \ M_\odot/pc^3$  and  $M_D(3.4 \pm 1.2) \times   10^7 \  M_\odot$  The uncertainties are reported for completeness they have no role in the results of this paper. 

Notice that  M 33 is  a clear case in which, although the best fit  $V_{IRCH}(R/R_{opt})$ is remarkably different  according to that  we force the halo density  to a URC-cored or a to a NFW-cuspy profile, $g_h$,  the  halo contribution to the radial acceleration,  takes similar values  in the two different cases.
  
\begin{figure}
\centering
\includegraphics[width=9.5cm]{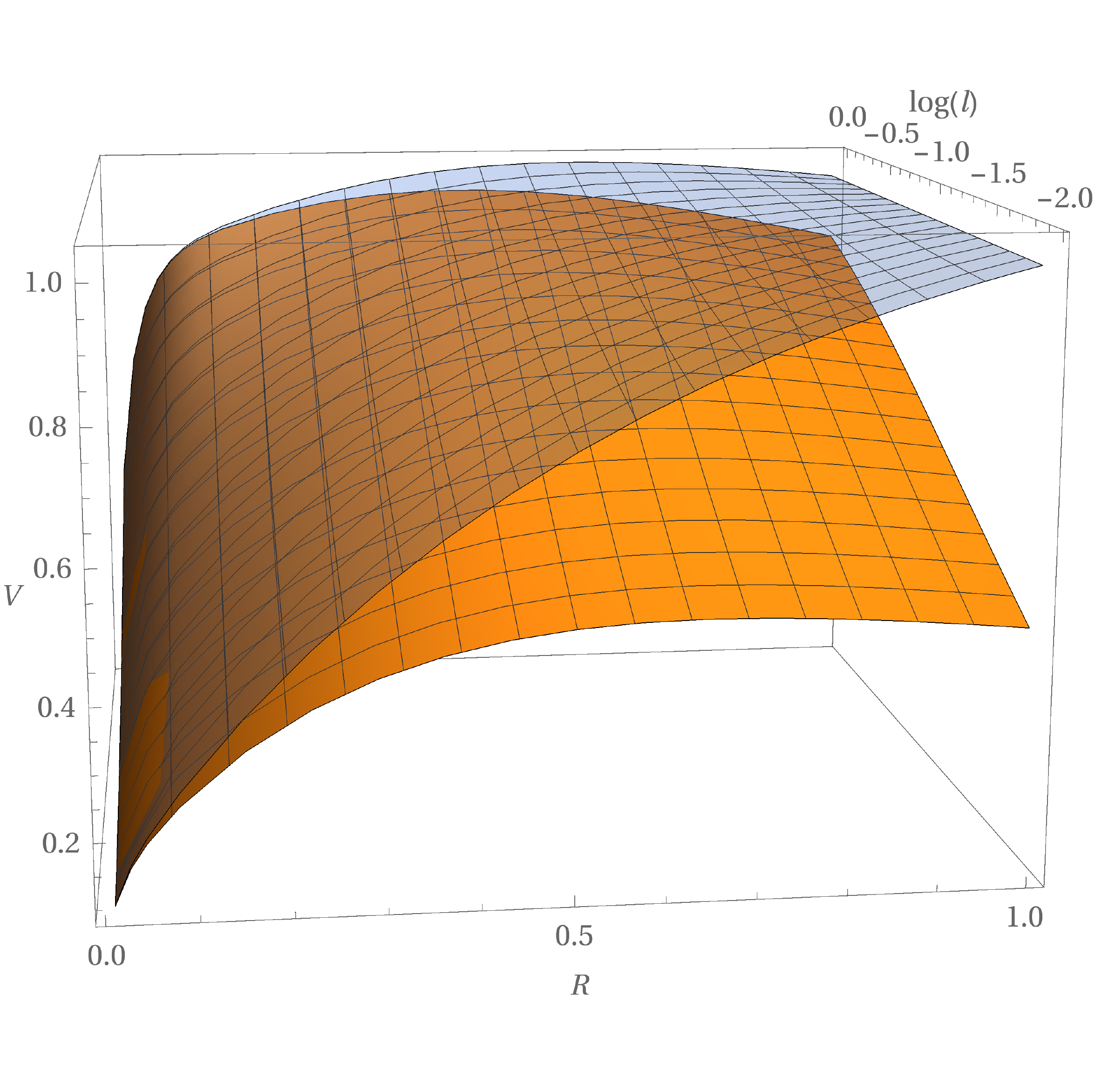}
\caption{ $V(x, l)/V(1, l)/$ and $V_b(x, l)/V(1, l) $ obtained by the RTF method.$ log \ l=  (M_I-23.5)/5$ }
\end{figure}

\vfill\eject
  \end{document}